\documentclass[10pt,twocolumn]{article}

\usepackage{graphicx} 
\graphicspath{{figures}} 
\usepackage[font=small,labelfont=bf]{caption} 

\usepackage{xcolor}
\definecolor{black}{HTML}{212427}
\definecolor{blue}{HTML}{0563C1}
\definecolor{brightred}{HTML}{FF0000} 

\usepackage{hyperref,nameref}
\hypersetup{colorlinks=true,allcolors=blue}
\newcommand{\rref}[2]{\hyperref[#1]{\ref{#1}#2}} 

\color{black}
\usepackage[margin=0.67in]{geometry} 
\usepackage[switch]{lineno} 

\usepackage{fancyhdr,lastpage} 
\usepackage{titlesec}
\titleformat{\section}{\bfseries}{}{1mm}{}
\titleformat{\subsection}{\bfseries}{\thesubsection}{1mm}{}
\titlespacing{\section}{0pt}{10pt}{0pt}
\titlespacing*{\section}{-3pt}{\baselineskip}{0pt}
\titlespacing{\subsection}{0pt}{10pt}{0pt}

\usepackage{amsmath,amssymb}
\newcommand{\Ang}[0]{\mathring{\mathrm{A}}} 
\let\f=\frac 
\renewcommand{\t}[1]{\text{#1}} 

\usepackage{multirow} 

\usepackage[
  backend=biber,        
  autocite=superscript, 
  sorting=none,         
  style=numeric-comp,   
  maxnames=100,         
  minnames=100,         
  giveninits=false,     
  autopunct=false,      
  doi=true,             
  hyperref=true         
]{biblatex}
\DeclareFieldFormat{labelnumberwidth}{\;[#1]\!\!\!\!} 
\renewbibmacro{in:}{} 
\AtNextBibliography{\small} 
\AtEveryBibitem{\clearfield{pages}} 
\AtEveryBibitem{\clearfield{volume}} 
\AtEveryBibitem{\clearfield{number}} 

\begin{document}
\twocolumn[
  \begin{center}
	\large
	\textbf{Capturing short-range order in high-entropy alloys with machine learning potentials}
  \end{center}

  Yifan Cao$^1$,
  Killian Sheriff$^1$, and
  Rodrigo Freitas$^1${\footnotemark[1]} \\
  $^1$\textit{\small Department of Materials Science and Engineering, Massachusetts Institute of Technology, Cambridge, MA, USA} \\

  {\small Dated: \today}

  \vspace{-0.15cm}
  \begin{center}
	\textbf{Abstract}
  \end{center}
  \vspace{-0.35cm}
  Chemical short-range order (SRO) affects the distribution of elements throughout the solid-solution phase of metallic alloys, thereby modifying the background against which microstructural evolution occurs. Investigating such chemistry-microstructure relationships requires atomistic models that act at the appropriate length scales while capturing the intricacies of chemical bonds leading to SRO. Here we consider various approaches for the construction of training data sets for machine learning potentials (MLPs) for CrCoNi and evaluate their performance in capturing SRO and its effects on materials quantities of relevance for mechanical properties, such as stacking-fault energy and phase stability. It is demonstrated that energy accuracy on test sets often does not correlate with accuracy in capturing material properties, which is fundamental in enabling large-scale atomistic simulations of metallic alloys with high physical fidelity. Based on this analysis we systematically derive design principles for the rational construction of MLPs that capture SRO in the crystal and liquid phases of alloys.
  \vspace{0.4cm}
]
{
  \footnotetext[1]{Corresponding author (\texttt{rodrigof@mit.edu}).}
}

\noindent In high-entropy alloys (HEAs)\autocite{hea_1,hea_2,hea_nrm} multiple metallic elements are combined in nearly equal concentrations. This often leads to the stabilization of crystalline phases in which elements are distributed throughout the alloy in a nearly-random fashion --- namely, solid solution phases. This class of alloys has attracted substantial interest due to their mechanical properties. For example, extraordinary fracture resistance was observed in CrCoNi\autocite{liu_exceptional_2022,gludovatz2016exceptional} as the result of an unusual synergy of deformation mechanisms involving stacking-fault formation and phase transitions, as well as the conventional gliding of dislocations. 

It has been established that chemical short-range order (SRO) --- i.e., the tendency of solid solutions to not be completely random --- affects various chemistry--microstructure relationships that influence mechanical properties. For example, SRO has been shown to affect dislocation mobility\autocite{Li2019, yin_atomistic_2021, chen2023short}, grain boundaries\autocite{megan_2,tim,penghui}, stacking-fault energy\autocite{PNAS_bob, zhang_short-range_2020, resistivity_easo, zhao2017stacking}, and phase stability\autocite{niu_magnetically-driven_2018}. Consequently, significant experimental efforts have been made to characterize SRO and its effects on materials properties\autocite{zhang_short-range_2020, xu_determination_2023, schonfeld_local_2019, inoue_direct_2021, chen_direct_2021, zhou_atomic-scale_2022, diffuse_origin, resistivity_easo, penghui}. Connecting computational results to such experiments requires high-fidelity physical models capable of capturing the intricate nature of chemical bonds leading to SRO, while also accounting for the complexity of chemical motifs in HEAs\autocite{first_paper, sheriff2024chemical}.

In a previous work (ref.~\cite{first_paper}) we have demonstrated that small-scale atomistic simulations, i.e., sizes typical of density-functional theory (DFT) calculations, are inadequate to properly capture SRO, leading to errors of up to 25\% in the prediction of Warren-Cowley parameters. In the same work an approach for training machine learning interatomic potentials (MLPs) was demonstrated to capture SRO while simultaneously leading to an improvement in energy accuracy when compared to the state-of-the-art. Yet, fundamentally, such an approach consisted of a set of heuristics on the construction of the MLP, i.e., a reasonable and practical approach for the construction of training sets without rigorous justification. Here we build on these results and systematically derive the design principles for the rational construction of MLPs that capture SRO.

While much of the work on MLPs for HEAs has focused on their energy accuracy over test data sets\autocite{shapeev_moment_2016, THOMPSON2015316, Behler2007, bartok_gaussian_2010, batzner_e3-equivariant_2022, musaelian_learning_2023,du_chemical_2022, ghosh_short-range_2022, kostiuchenko_impact_2019, kostiuchenko_short-range_2020, li_complex_2020, yin_atomistic_2021, zheng_multi-scale_2023, yu_theory_2022}, here we focus instead on their performance in reproducing SRO and its effects on materials quantities of relevance for mechanical properties. We systematically augment an initially simple MLP training set while tracking the associated effects on the SRO of the crystal and liquid phases, stacking-fault energy, and phase stability. It is demonstrated that energy accuracy on test data sets often does not correlate with accuracy in capturing such material properties, which are fundamental in enabling large-scale atomistic simulations of HEAs with high physical fidelity.

\begin{figure*}[!htb]
  \centering
  \includegraphics[width=\textwidth]{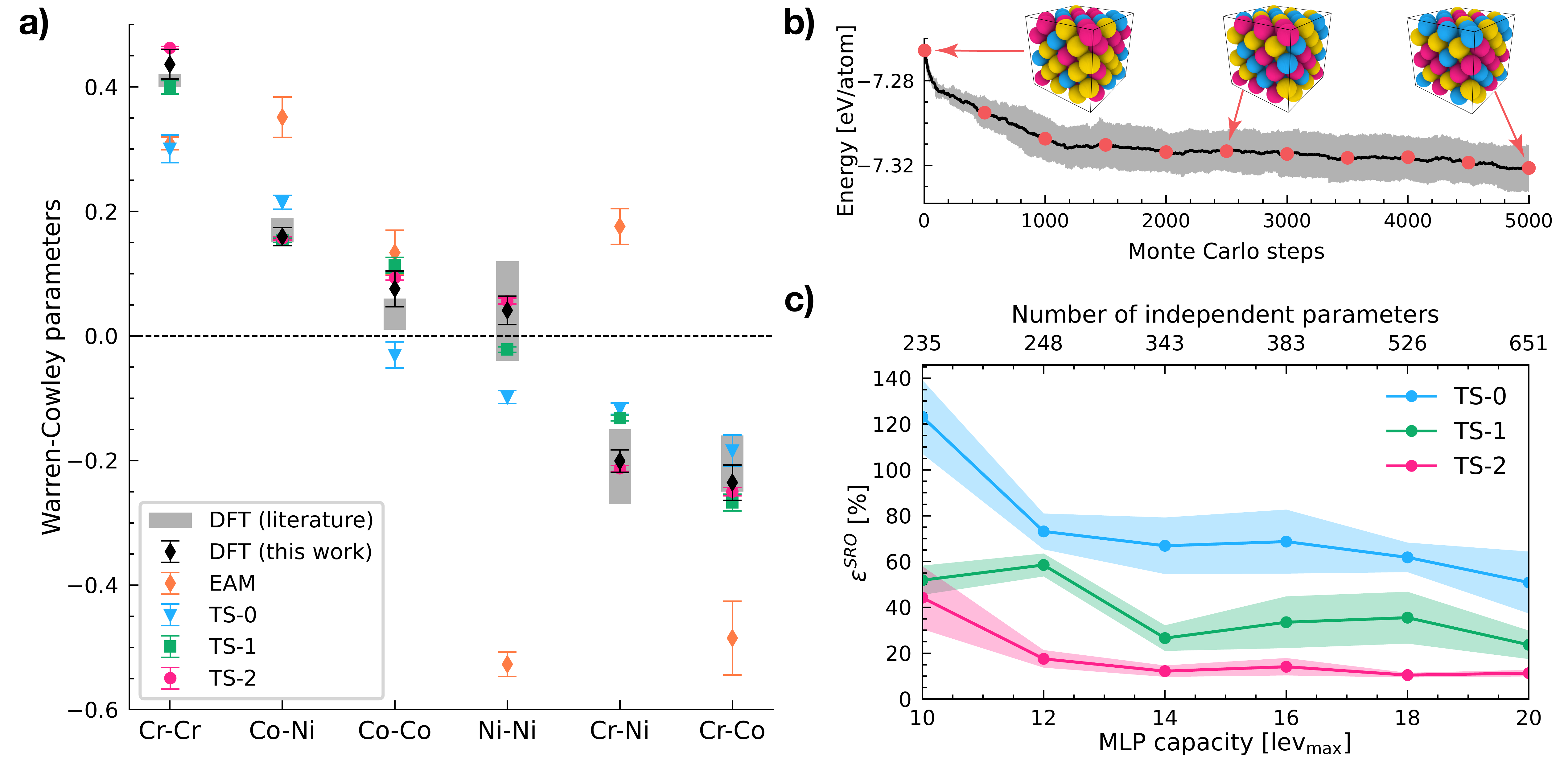}
  \caption{\label{figure_1} \textbf{Capturing chemical SRO in the crystal phase.} \textbf{(a)} Comparison of predicted WC parameters against DFT Monte Carlo (literature values from refs.~\cite{PNAS_bob} and \cite{tamm_atomic-scale_2015}). The EAM potential is from ref.~\cite{Li2019} while the training sets for potentials TS-0, TS-1, and TS-2 are summarized in table~\ref{table_1}. Error bars are the standard error from the mean from an ensemble of 20 independent potentials. \textbf{(b)} Illustration of the intermediary configurations extracted from DFT Monte Carlo to train TS-2. \textbf{(c)} Relative error with respect to DFT (eq.~\ref{eq:SRO_error}) as a function of the MLP model capacity. Error bars are the 95\% confidence interval from an ensemble of 20 independent potentials.}
\end{figure*}

\section{Results}
\section{Training strategy to manage chemical complexity}
In order to develop the design principles for capturing SRO we focus on the face-centered cubic (fcc) solid-solution phase of the paradigmatic CrCoNi alloy\autocite{liu_exceptional_2022, gludovatz2016exceptional, SRO_radiation,PNAS_bob,zhang_short-range_2020,flynn_PNAS, LCO_PRL,tamm_atomic-scale_2015,Li2019,zhou_atomic-scale_2022, resistivity_easo}. The MLP model chosen is the Moment Tensor Potential\autocite{shapeev_moment_2016} with a radial cutoff of $5\,\Ang$, which corresponds to a distance in between the 3rd and 4th coordination shell of CrCoNi. In the absence of any other criteria to guide our initial choice of ML model, we have chosen to employ MTP due to its superior performance in energy and force errors for single-element systems compared to other MLP models (as demonstrated by an independent assessment in ref.~\cite{zuoPerformanceCostAssessment2020}). An a posteriori analysis described in Supplementary Section 1 shows that MTP also has superior performance for various material properties of the CrCoNi alloy when compared to a few other ML models.

In ref.~\cite{first_paper} we demonstrated that the first coordination shell in CrCoNi can be chemically decorated in 36,333 unique configurations (i.e., chemical motifs). The relative energy of these chemical motifs affects the frequency with which they are observed in the alloy --- which is the fundamental origin of SRO --- making this an important property to be reproduced by MLPs in order to capture SRO. Yet, while the first coordination shell is the dominating term in atomic interactions, many-body contributions from higher coordination shells are not negligible and must be accounted in the development of MLPs\autocite{Behler2007, behler2011neural, bartok2013representing, THOMPSON2015316, shapeev_moment_2016, drautz2019atomic}. Extending the counting of unique chemical motifs to the second and third coordination shells results in approximately $2.5 \times 10^7$ and $6.8 \times 10^{18}$ unique motifs respectively. Comparing these numbers with the 651 independent parameters of the MLP model with highest capacity employed here makes it clear that chemical complexity of this magnitude leads to a landscape with many nearly-degenerate minima for the MLP fitting. Here this chemical complexity is considered by employing an ensemble training approach\autocite{first_paper, ghosh_short-range_2022, kostiuchenko_short-range_2020}: multiple potentials are fitted under identical training conditions. Variations in performance among the resulting potentials due to the nearly-degenerate minima landscape are explicitly evaluated against materials properties related to SRO, including the role of the model capacity (i.e., number of independent parameters).

In the following sections we gradually build towards a final training set by systematically augmenting an initially simple training set while evaluating the effect of the modifications on associated material properties. In order to focus on the role of chemical complexity we employ a standard approach for accounting for thermal vibrations and thermal expansion in all potentials (described in the Methods section). Throughout this process the performance is also compared to a benchmark training set referred to as ``TS-0'', which was first introduced in ref.~\cite{first_paper} as ``training set without chemical sampling''. This training set was built by adapting the popular approach introduced in ref.~\cite{li_complex_2020} for bcc NbMoTaW to CrCoNi, which includes perfect and distorted ground state structures, slab structures, and molecular dynamics structures spanning single element, binary, and ternary element systems. While the approach in ref.~\cite{li_complex_2020} was not developed with the intent of capturing SRO, it is one of the most comprehensive and popular approaches for constructing training sets for HEAs\autocite{zheng_multi-scale_2023, yin_atomistic_2021}, which warranted its choice as benchmark data set.

\section{Chemical SRO in the crystal phase}
Quantification of SRO in the crystalline phase can be performed by evaluating the Warren-Cowley (WC) parameters:
\begin{equation}
  \label{eq:wc}
  \alpha_{ij} = 1 - \f{p(i|j)}{c_i},
\end{equation}
where $i$ and $j$ refer to any of the three chemical elements in the alloy, $c_i$ is the average concentration of $i$-type atoms, and $p(i|j)$ is the conditional probability of finding a $i$-type atom in the first coordination shell of an $j$-type atom. The effectiveness of MLPs in capturing SRO will be quantified by comparing WC parameters against those obtained through DFT Monte Carlo simulations. In fig.~\rref{figure_1}{a} we show that a popular interatomic potential\autocite{Li2019} (embedded-atom model, or EAM) for CrCoNi is not capable of reproducing WC parameters. Similarly, TS-0 also falls short of reproducing DFT results within the statistical accuracy, despite resulting in a considerable improvement in comparison with EAM.

We turn now to the construction of our initial training set, named TS-1. This training set is composed of chemically random and equiatomic fcc supercells with 108 atoms, adding up to 54,540 atoms (as summarized in table~\ref{table_1}). Despite its simplicity, it is clear from fig.~\rref{figure_1}{a} that TS-1 outperforms TS-0 for all WC parameters, which can only be attributed to the more extensive sampling of the chemical space performed in TS-1 when compared to TS-0. Despite this encouraging result, note once again that the chemical space is enormous compared to the number of independent parameters in the MLP model. Thus, it is reasonable to expect that sampling the chemical space with an approach better than random will lead to improved performance. We propose to accomplish this by substituting the chemically random configurations of TS-1 with intermediary configurations from DFT Monte Carlo simulations, as illustrated in fig.~\rref{figure_1}{b}. We name this training set TS-2. The configurations along the Monte Carlo trajectory exhibit an increasing amount of SRO with small energetic differences among them, all associated with changes in local chemical motifs. They function as a guide for the MLP, nudging it to capture the most relevant regions of the chemical space (i.e., regions that show up frequently due to SRO, as shown in refs.~\cite{first_paper} and \cite{megan}), as well as the trajectory to arrive at SRO from an initially random solid solution. It can be seen in fig.~\rref{figure_1}{a} that TS-2 is able to reproduce all WC parameters predicted by DFT within the statistical accuracy.

The performance of each MLP in fig.~\rref{figure_1}{a} can be summarized by evaluating the relative error with respect to DFT:
\begin{equation}
  \label{eq:SRO_error}
  \varepsilon^\t{SRO} = \f{\sum\limits_{i=1}^3 \sum\limits_{j=i}^3 \left | \alpha_{ij}^\t{MLP}-\alpha_{ij}^\t{DFT} \right |}{\sum\limits_{i=1}^3 \sum\limits_{j=i}^3 \left | \alpha_{ij}^\t{DFT} \right |}.
\end{equation}
This relative error is shown in fig.~\rref{figure_1}{c} for all three training sets as a function of the MLP model capacity (i.e., number of independent parameters), where it can be seen that the observations above regarding the superior performance of TS-2 relative to TS-0 and TS-1 hold for any model capacity. More importantly, note how TS-2 has significantly more stable behavior against random weight initialization for training, which is shown in fig.~\rref{figure_1}{c} by an ensemble standard deviation of $3\%$ at lev$_\t{max} = 20$ compared to $29\%$ for TS-0 and $13\%$ for TS-1. Given the simplicity and similarity of the training sets for TS-1 and TS-2, this smaller variation within the ensemble can only be attributed to the extensive chemical sampling and the targeted sampling of motifs of relevance for SRO.

\begin{table}[tb]
  \centering
  \begin{tabular}{c | c c | c c c | c}
    \hline \hline
    \multirow{2}{*}{\textbf{Name}} & \multicolumn{2}{c|}{\textbf{Chemical}} & \multicolumn{3}{c|}{\textbf{Phase fraction}} & \textbf{Size}    \\
                                   & \multicolumn{2}{c|}{\textbf{order}}    & \multicolumn{3}{c|}{\textbf{(\%)}}           & \textbf{(atoms)} \\
    \hline
                                   & RSS & SRO                              & fcc & hcp & liquid                                            & \\
    \hline
    TS-0\footnotemark[2] & yes & no  & 68  & 0  & 32 & 83,970 \\
    TS-1 & yes & no  & 100 & 0  & 0  & 54,540 \\
    TS-2 & yes & yes & 100 & 0  & 0  & 54,540 \\
    \hline
    TS-3\footnotemark[3] & yes & yes & 50  & 0  & 50 & 108,684 \\
    \hline
    TS-4 & yes & yes & 100 & 0  & 0  & 36,720 \\
    TS-5 & yes & yes & 50  & 50 & 0  & 73,440 \\
    \hline
    TS-f & yes & yes & 25  & 25 & 50 & 146,880 \\
    \hline \hline
    \multicolumn{7}{l}{\small \footnotemark[2] Same training set as ``without chemical sampling'' in ref.~\cite{first_paper}.} \\
    \multicolumn{7}{l}{\small \footnotemark[3] Same training set as ``with chemical sampling'' in ref.~\cite{first_paper}.}
  \end{tabular}
  \caption{\label{table_1} \textbf{Summary of the contents of each MLP training set.} If the data set includes at least one random solid solution (RSS) configuration it is marked as ``yes'' in that column. The chemical order ``SRO'' column indicates that configurations extracted from DFT Monte Carlo were included. The training set for TS-0 was motivated by ref.~\cite{li_complex_2020}.}
\end{table}

\section{Liquid stability and SRO}
\begin{figure*}[htb!]
  \centering
  \includegraphics[width=\textwidth]{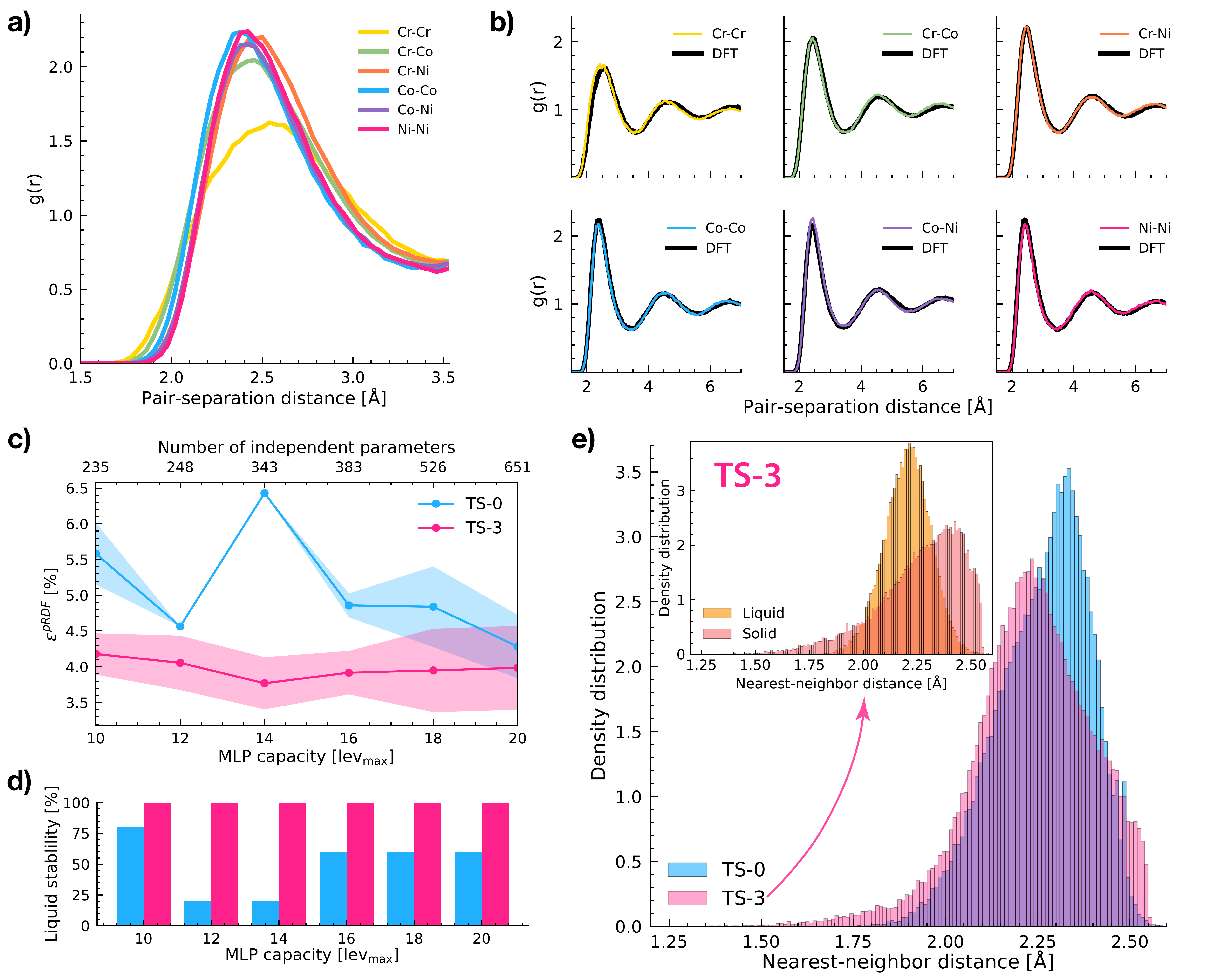}
  \caption{\label{figure_2} \textbf{Capturing chemical and structural SRO in the liquid phase.} \textbf{(a)} Partial radial distribution functions (pRDF) of the liquid phase at 2684\,K from DFT. \textbf{(b)} Comparison of pRDF using TS-3 (colored lines) against DFT values (black lines). \textbf{(c)} Relative error with respect to DFT (eq.~\ref{eq:pRDF_error}) in the prediction of structural and chemical SRO in the liquid as a function of MLP model capacity. \textbf{(d)} Fraction of potentials in the ensemble with stable liquid phase. Blue bars are for TS-0 and pink bars are for TS-3. \textbf{(e)} Distribution of nearest-neighbor distance for each atom in TS-0 and TS-3. Inset shows the breakdown of nearest-neighbor distance in TS-3 by phase. The success of TS-3 in reproducing a stable liquid phase is attributed to the inclusion of crystal configurations with atom pairs at much closer distances than TS-0.}
\end{figure*}

The liquid phase is also afflicted by the considerable chemical complexity observed in the crystal phase. The concept of SRO becomes even more complex in the liquid because chemical and structural SRO are both present. This can be observed in the partial radial distribution functions (pRDFs) obtained from DFT molecular dynamics simulations at 2684\,K shown in fig.~\rref{figure_2}{a}: the Cr-Cr and Cr-Co peaks are notably lower than other peaks, indicating that these chemical pairs are energetically unfavorable (i.e., chemical SRO). An entanglement between chemical and structural SRO leads to the variations of the first coordination shells in fig.~\rref{figure_2}{a}.

The effects observed in fig.~\rref{figure_2}{a} are addressed by augmenting TS-2 with configurations obtained from DFT molecular dynamics simulations of the liquid phase at 1800\,K. The resulting training set --- named TS-3 and summarized in table~\ref{table_1} --- is able to capture the chemical and structural SRO of the pRDFs, as shown in fig.~\rref{figure_2}{b}: the first coordination shell peak heights are in agreement within $\pm 4.9\%$, while the major discrepancy in the first coordination shell peak location being $-0.12\,\Ang$ for Cr-Cr pairs.

\begin{figure*}[htb]
  \centering
  \includegraphics[width=\textwidth]{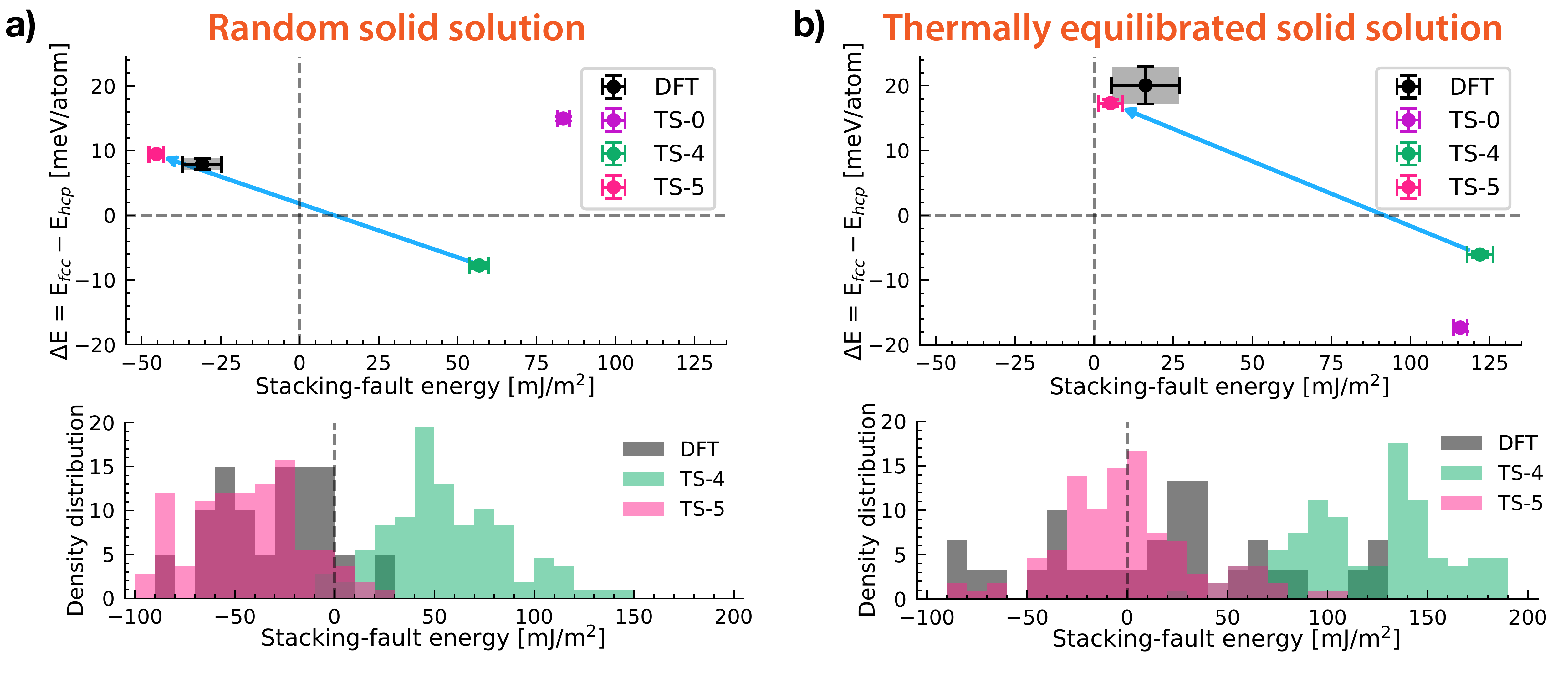}
  \caption{\label{figure_3} \textbf{Capturing SRO effects on (111) stacking-fault energy ($\gamma_\t{sf}$) and fcc--hcp phase stability ($\Delta E = E_\t{fcc}-E_\t{hcp}$).} Comparison of $\gamma_\t{sf}$ and $\Delta E$ against DFT results for \textbf{(a)} random solid solutions and \textbf{(b)} solid solutions in thermal equilibrium (i.e., with appropriate SRO as obtained through Monte Carlo simulations). The blue arrow indicates the improvements in both quantities by augmenting TS-4 with hcp configurations. Error bars are the standard error from the mean from independent Monte Carlo simulations. Figures at the bottom indicate the density distribution of $\gamma_\t{sf}$ due to chemical fluctuations and SRO.}
\end{figure*}

The performance in capturing chemical and structural SRO in the liquid can be summarized by evaluating the relative error in the absolute differences between the pRDFs with respect to DFT, similarly to what was done for the WC parameters in eq.~\ref{eq:SRO_error}:
\begin{equation}
  \label{eq:pRDF_error}
  \varepsilon^\t{pRDF} = \f{\sum\limits_{i=1}^3 \sum\limits_{j=i}^3 \displaystyle\int_{0}^{r_\t{max}} \left | g_{ij}^\t{MLP}(r)-g_{ij}^\t{DFT}(r) \right | \t{d}r}{\sum\limits_{i=1}^3 \sum\limits_{j=i}^3 \displaystyle\int_{0}^{r_\t{max}} \left | g_{ij}^\t{DFT}(r) \right | \t{d}r} ,
\end{equation}
where $g_{ij}(r)$ is the pRDF between chemical elements $i$ and $j$, and $r_\t{max} = 3.5\,\Ang$ is the extension of the first coordination shell (i.e., total RDF minimum between first and second peaks), which was chosen as to evaluate only the short-range part of chemical and structural ordering. Using this approach one can see in fig.~\rref{figure_2}{c} that TS-3 has better performance than TS-0, which is unexpected because TS-0 is trained with configurations extracted from the same simulation employed to collect the pRDF statistics in fig.~\rref{figure_2}{b} (i.e., at 2684\,K) while TS-3 configurations were collected at 1800\,K. Notice that one is unable to detect this performance difference in reproducing materials properties by evaluating only the energy root-mean-square error: TS-0 and TS-3 result in 5.3\,meV/atom and 5.6\,meV/atom respectively at 1800\,K, and 7.6\,meV/atom and 6.6\,meV/atom at 2684\,K.

Another important consequence of chemical complexity is the fact that a large fraction of the ensemble of MLPs obtained with TS-0 were unstable in the liquid phase, i.e., simulations of the liquid phase with these potentials quickly encountered configurations with unphysically large values for forces and energies that led the simulation to fail. Notice that this behavior is observed despite the inclusion of liquid configurations in the training set of TS-0 potentials. The fraction of potentials in the ensemble with unstable liquid phase is shown in fig.~\rref{figure_2}{d} as a function of the model capacity. Note how for lev$_\t{max} = 12$ and $14$ only one out of the five potentials trained is stable. In the same figure it can be seen that TS-3 never results in such instabilities.

We were able to track down the source of TS-3 success in stabilizing the liquid to the inclusion of crystal configurations with atomic pair distances much closer together than those observed in TS-0, as can be seen in the histogram of fig.~\rref{figure_2}{e} (see Supplementary Section 2 for a detailed breakdown of this analysis). This observation is explained by further considering the role of chemical complexity as follows. Despite its disordered structure, the liquid has well-defined structural SRO\autocite{di2003there, jonsson1988icosahedral} (e.g., the three coordination shells clearly visible in fig.~\rref{figure_2}{b}) leading to a similar degree of chemical complexity as the crystalline phase\autocite{first_paper}. Yet, differently from the crystalline phase, atoms in the liquid are frequently diffusing around, which requires going through activated transition states with higher potential energy due to the close proximity of atoms\autocite{iio,sam}. Such transition states are considered ``rare events'' in the course of a simulation such as the ones used to extract configurations for TS-0 and TS-3 because they occur only every many timesteps for each atom. It is suspected that the inclusion of crystal configuration with close pair distances (fig.~\rref{figure_2}{e}) in TS-3 induces the learning of the energetics of such transition states in between chemical motifs, rendering the liquid phase stable.

\section{Stacking-fault energy and phase stability}
Mechanical properties of fcc alloys are closely linked to the $(111)$ stacking-fault energy ($\gamma_\t{sf}$) and the relative stability of the fcc phase with respect to the hexagonal close-packed (hcp) phase\autocite{zhang_dislocation_2017,miao_evolution_2017, laplanche2017reasons} ($\Delta E = E_\t{fcc} - E_\t{hcp}$). For example, engineering of phase metastability with $\Delta E$ enables transformation-induced plasticity\autocite{li_metastable_2016}, while lowering $\gamma_\t{sf}$ is associated with a change in plasticity mechanism from dislocation slip to twinning. Notably, the exceptional damage tolerance\autocite{liu_exceptional_2022} of CrCoNi is the result of an unusual synergy between the deformation mechanisms described above.

Figure \ref{figure_3} shows that TS-0 is not capable of reproducing $\gamma_\t{sf}$ or $\Delta E$. Our strategy for capturing these two quantities with MLPs is centered around the well-established correlation between them: a fcc-to-hcp transition can be accomplished by the successive introduction of stacking-faults on every other $(111)$ plane. Thus, it is reasonable to expect that training with configurations of one of these structures (i.e., stacking faults or hcp) is enough to learn the other. We chose to employ hcp configurations instead of stacking faults because they are straightforward to simulate. Two training sets were created to compare the performance of MLPs trained with and without hcp configurations. The first training set (TS-4) includes only fcc configurations and is similar to TS-2 in all aspects except for its total size, which is reduced by 32.7\% with respect to TS-2 (as shown in table~\ref{table_1}). This reduction is performed with the goal of accommodating an equivalent amount of hcp configurations in the second training set (TS-5), thereby achieving parity between fcc and hcp configurations while maintaining a moderate training set total size. With TS-4 and TS-5 we seek to establish the importance of hcp configurations in capturing $\gamma_\t{sf}$ and $\Delta E$; the liquid phase was intentionally left out of both training sets (i.e., TS-4 and TS-5) to avoid interference with this test.

\begin{figure}[htb]
  \centering
  \includegraphics[width=0.5\textwidth]{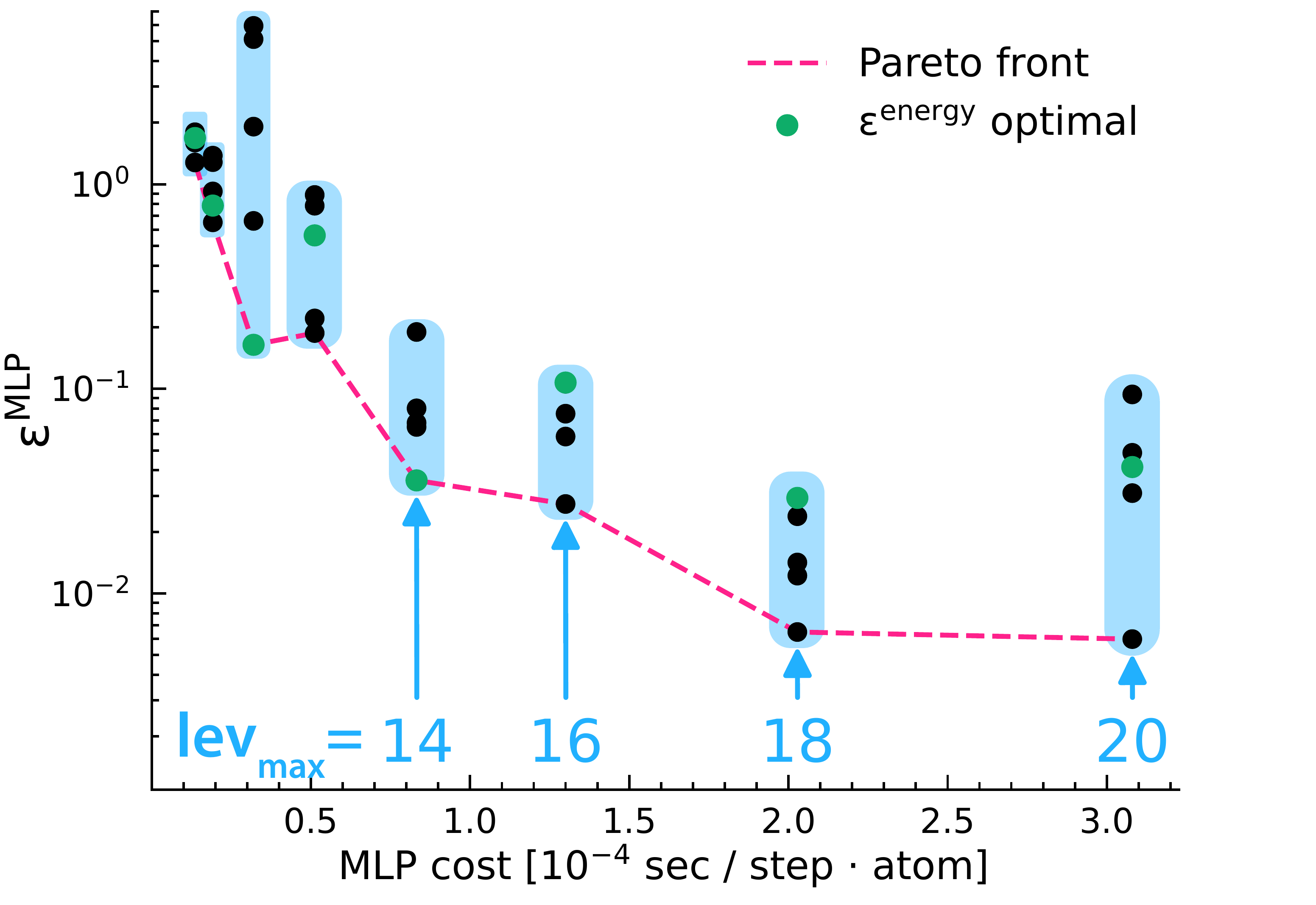}
  \caption{\label{figure_4} \textbf{Final potential (TS-f) performance across materials properties.} The best performing potential for a fixed computational cost falls on the Pareto front. Blue rectangles mark ensembles of identically-trained potentials for different model capacities (i.e., lev$_\t{max}$). Potentials with lowest root-mean-square energy error ($\varepsilon^\t{energy}$) within each ensemble are marked in green. The four ensembles on the left have lev$_\t{max} = 6$, 8, 10, and 12.}
\end{figure}

Figure \rref{figure_3}{a} compares $\gamma_\t{sf}$ and $\Delta E$ against DFT for a random solid solution at 500\,K, where it can be seen that the introduction of the hcp phase is fundamental in capturing the correct value of $\gamma_\t{sf}$ and $\Delta E$. Yet, SRO in CrCoNi leads to an increase in $\gamma_\t{sf}$\autocite{PNAS_bob}, thus we also compare $\gamma_\t{sf}$ and $\Delta E$ for a solid solution in thermal equilibrium at 500\,K in fig.~\rref{figure_3}{b}. While the agreement in fig.~\rref{figure_3}{b} might seem obvious in light of fig.~\rref{figure_3}{a}, we warn that such anticipation is not warranted. Figure ~\rref{figure_3}{b} is a much more stringent test than simply capturing the energetics of stacking faults and crystalline phases, which MLPs are well-known to be capable of capturing\autocite{freitas_machine-learning_2022, pun_development_2020, fujii_structure_2022, bartok_machine_2018, goryaeva_efficient_2021}. Instead, fig.~\rref{figure_3}{b} demonstrates that TS-5 is capable of capturing the correct SRO \textit{and} its effects on materials properties (i.e., $\gamma_\t{sf}$ and $\Delta E$). This is because independent Monte Carlo simulations for thermal equilibration were performed in each case (DFT, TS-4, and TS-5), leading to independent predictions of SRO configurations, which were then employed to evaluate $\gamma_\t{sf}$ and $\Delta E$.

\begin{figure*}[htb]
  \centering
  \includegraphics[width=\textwidth]{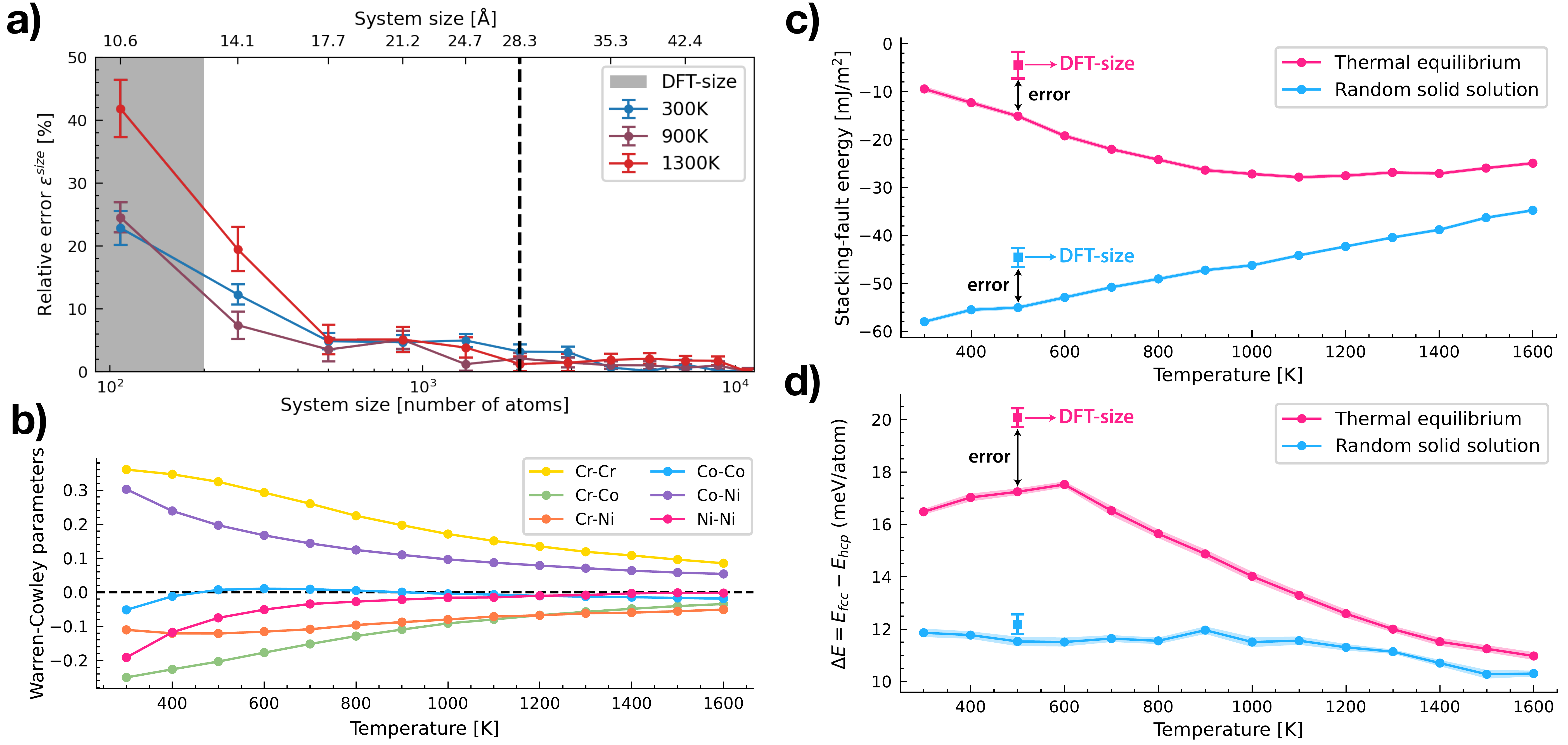}
  \caption{\label{figure_5} \textbf{Final potential (TS-f) prediction of material properties with size-converged SRO.} \textbf{(a)} Relative error of WC parameters with system size ($\varepsilon^\t{size}$), measured similarly to eq.~\ref{eq:SRO_error} but against the largest size prediction. Simulations with less than 2,000 atoms are not converged with respect to system size. \textbf{(b)} Temperature dependence of WC parameters (eq.~\ref{eq:wc}). \textbf{(c)} Temperature dependence of stacking-fault energy ($\gamma_\t{sf}$) for a random solid solution and a solid solution in thermal equilibrium (i.e., with appropriate SRO as obtained through Monte Carlo simulations). The ``DFT-size'' data point was evaluated with TS-f, but using a system with only 180 atoms, which is the typical size of DFT calculations (as indicated in fig.~\rref{figure_5}{a}). \textbf{(d)} Temperature dependence of the fcc--hcp phase stability. See text for a full discussion of the effect of SRO and the phase transition at low temperatures leading to the kink at 600\,\t{K} for solid solutions in thermal equilibrium.}
\end{figure*}

\section{Final training set}
A final training set (TS-f) is constructed (see table \ref{table_1}) to incorporate all elements leading to good performance on capturing SRO and its effects on the crystal phase, liquid phase, stacking faults, and phase stability. Its performance across the various materials properties is evaluated through the following metric:
\begin{equation}
  \label{eq:Pareto_optimality_scheme}
  \varepsilon^\t{MLP} = \varepsilon^\t{SRO}_\t{fcc} \times \varepsilon^\t{SRO}_\t{hcp} \times \varepsilon_\t{liquid}^\t{pRDF} \times \varepsilon^\t{energy} ,
\end{equation}
where $\varepsilon^\t{SRO}_\t{fcc}$ and $\varepsilon^\t{SRO}_\t{hcp}$ are the relative WC errors (eq.~\ref{eq:SRO_error}) for the fcc and hcp phases, $\varepsilon_\t{liquid}^\t{pRDF}$ is shown in eq.~\ref{eq:pRDF_error}, and $\varepsilon^\t{energy}$ is the energy root-mean-square error over a test set with a wide range of configurations, including random solid phases, thermally equilibrated solid solutions, and liquid phases (see methods for a full description). The $\varepsilon^\t{MLP}$ error as a function of the computational cost for different model capacities is shown in fig.~\ref{figure_4} along with the Pareto front. The large variation in $\varepsilon^\t{MLP}$ performance within an ensemble with same model capacity (lev$_\t{max}$) is noteworthy (note the log-scale in the y-axis). For example, the best performing MLP with lev$_\t{max} = 14$ has $\varepsilon^\t{MLP}$ comparable to the second best out of the five MLPs in the ensemble with lev$_\t{max} = 20$, despite the latter being $\approx3.6$ times more computationally expensive. Moreover, note how the potential with lowest $\varepsilon^\t{energy}$ is often not the best performing potential for materials properties: $\varepsilon^\t{energy}$ only predicts the best performance in two out of the eight ensembles in fig.~\ref{figure_4}.

\section{Material properties with size-converged SRO}
Equipped with TS-f we now turn to demonstrate the effect of SRO on properties and scales that cannot be achieved in the absence of the MLP introduced here. We start by reproducing the observation, made in ref.~\cite{first_paper}, that DFT-sized calculations are not converged with respect to size and produce relative errors up to 40\% in SRO because the SRO characteristic length scale can be as large as $25\,\Ang$ to $30\,\Ang$. This can be seen in fig.~\rref{figure_5}{a}, where the relative error of WC parameters with respect to a large calculation with 10,976 atoms is shown as a function of system size. Systems with as many as 2,000 atoms (or, equivalently, linear dimensions of $28\,\Ang$) are required for size convergence, while DFT calculations for SRO investigations in the literature use one order of magnitude less atoms\autocite{PNAS_bob, tamm_atomic-scale_2015}. We further demonstrate here that similar size convergence is necessary for predicting material properties strongly influenced by SRO. As illustrated in figs.~\rref{figure_5}{c} and \rref{figure_5}{d}, DFT-size simulations yield significant errors of $10.7\,\t{mJ/m}^2$ in stacking fault energy ($\gamma_{SF}$) and $2.84\,\t{meV/atom}$ in phase stability ($\Delta E$) under thermal equilibrium (i.e., with appropriate SRO as obtained through Monte Carlo simulations).

The temperature dependence of the WC parameters was evaluated with calculations converged with respect to system size. Figure \rref{figure_5}{b} shows that in general the WC parameters decrease in magnitude as the temperature increases, but this behavior is not monotonous for $\alpha_\t{CoCo}$, where an increase in magnitude with temperature is observed above 900\,K. This is yet another evidence of the incompleteness of WC parameters in quantifying SRO: in ref.~\cite{first_paper} we demonstrate that an appropriate and complete SRO metric shows smooth monotonic decrease in SRO with temperature.

\begin{table}[b]
  \centering
  \begin{tabular}{l c}
    \hline \hline
    \multicolumn{2}{c}{\textbf{Melting temperature (K)}} \\
    \hline
    Experimental\autocite{exp_melting_pt} & 1690 \\
    TS-f & 1661 \\
    EAM\autocite{Li2019} & 1410 \\
    \hline \hline
  \end{tabular}
  \caption{\label{table_2} \textbf{Comparison of melting temperature.} All computational results were obtained using the phase-coexistence method\autocite{morris1994melting}.}
\end{table}

The stacking-fault energy ($\gamma_\t{sf}$) dependence on temperature is also not trivial, as shown in fig.~\rref{figure_5}{c}. Short-range order increases $\gamma_\t{sf}$ at all temperatures with respect to a random solid solution, which is aligned with previous results\autocite{PNAS_bob}. This observation can be rationalized by the fact that stacking-fault creation in a thermally equilibrated solid solution requires the disruption of chemical motifs with lower energy than those encountered in a random solid solution. Yet, while the temperature effect on random solid solutions is a simple linear increase in $\gamma_\t{sf}$ (due to thermal expansion), a solid solution in thermal equilibrium (i.e., with appropriate SRO as obtained through Monte Carlo simulations) presents a complicated interplay between SRO and temperature. At low temperatures the contribution from SRO dominates and $\gamma_\t{sf}$ decreases almost linearly with temperature, but as temperature increases above around 800\,K $\gamma_\t{sf}$ displays a linear increase with temperature almost in parallel with the random solid solution. To our knowledge this is the first report of the complex temperature dependence of $\gamma_\t{sf}$, despite its central role in rationalizing many of the mechanical properties of CrCoNi and other fcc HEAs.

The fcc-hcp relative phase stability ($\Delta E$) is also evaluated with TS-f, as shown in fig.~\rref{figure_5}{d}. In general the presence of SRO further stabilizes hcp phase at all temperatures, i.e., increases $\Delta E$ with respect to the random solid solution. Yet, the absolute value of $\Delta E$ decreases  with increasing temperature, in agreement with previous first-principles calculations and thermodynamic models\autocite{saunders1998calphad, liu2009first} indicating that fcc phase becomes the stable phase near the melting temperature\autocite{niu_magnetically-driven_2018}. At low temperatures $\Delta E$ has a kink near 600\,K for solid solutions in thermal equilibrium that is not present in random solid solutions, which is consistent with previous findings\autocite{ghosh_short-range_2022,flynn_MRS_comm,du_chemical_2022} indicating a potential phase transition to an ordered phase at low temperatures.

Finally, the melting temperature of TS-f was evaluated using phase-coexistence molecular dynamics simulations. The agreement with experimental results is excellent, as shown in table \ref{table_2}, and a marked improvement over existing interatomic potentials\autocite{Li2019}. 

\section{Discussion and conclusion}
The most important design principle supported by the results presented here is that chemical complexity should be sampled extensively and biased towards chemical motifs of relevance for SRO. As shown in fig.~\rref{figure_1}{c}, this is the most effective strategy to reduce the impact of the nearly-degenerate landscape of minima on MLP fitting, which all other principles rely on to perform well in capturing materials properties. A promising venue to amplify the benefits of this design principle will be its combination with compressed lower-dimensional descriptors of chemical information\autocite{m1,m2,m3,m4_rss}, which were developed to alleviate the dramatic increase in MLP capacity required to accommodate an increasing number of chemical species.

Quantitative experimental characterization of SRO has not yet been achieved\autocite{flynn_nat_mat, diffuse_origin,EXAFS_SRO, resistivity_easo, flynn_diffraction_origin}, and the feasibility of employing SRO as a design feature for materials properties remains uncertain. In light of these observations it is justifiable to question the relevance of SRO in the field of HEAs. One argument for its relevance is the potential ubiquity of SRO in various materials properties and phenomena, as highlighted in ref.~\cite{flynn_MRS_comm} by historical results in metallurgy and a review of the fundamental principles of clustering in simpler alloys. The results presented here introduce another argument for the relevance of SRO: the importance of chemical complexity in the development of high-fidelity atomistic physical models for solid solutions, as exemplified by the following design principles: the liquid phase is only rendered stable (fig.~\ref{figure_2}) after careful consideration of its structural and chemical SRO and the inclusion of solid configurations with close atomic pair distances, which ought to also affect the solidification process (table \ref{table_2}) and as-cast state of these alloys. Similarly, the stacking-fault energy and phase stability (fig.~\ref{figure_3}) are only captured within DFT accuracy after accounting for chemical SRO equally in both fcc and hcp phases. 

The results obtained here highlight the importance of modeling SRO at the appropriate length scales. The observation that DFT-sized calculations are not converged with respect to SRO (i.e., fig.~\rref{figure_5}{a}) was first made in ref.~\cite{first_paper}. Yet, here we note that size convergence with respect to SRO is also important for properties such as stacking-fault energy (fig.~\rref{figure_5}{c}) and fcc-hcp relative phase stability (fig.~\rref{figure_5}{d}). Large-scale simulations dramatically reduce the uncertainty in the estimation of $\gamma_\t{sf}$ --- which has its origins in the chemical complexity (fig.~\ref{figure_3}) --- and reveal in fig.~\rref{figure_5}{c} that $\gamma_\t{sf}$ is negative at all temperatures. This observation supports the argument proposed in ref.~\cite{Maryam_stacking_fault} that positive stacking-fault \textit{free} energy is the likely explanation for finite partial-dislocation separations observed experimentally.

In conclusion, our work introduces a series of design principles for the construction of training data sets that optimize MLP performance in capturing SRO and its effects on important materials properties, such as phase stability and defect energies. The effectiveness of each design principle is confirmed by comparing the performance of MLPs trained with and without the proposed approaches in reproducing associated materials properties. A final training set that includes all proposed principles is produced and its performance across various materials properties is evaluated and summarized in a single metric (eq.~\ref{eq:Pareto_optimality_scheme}), thereby enabling large-scale atomistic simulations of CrCoNi and other HEAs with high physical fidelity.

\printbibliography[heading=bibnumbered,title={References}]

\end{document}